\documentclass[aps,pra,superscriptaddress,twocolumn,showpacs]{revtex4}

\renewcommand*{\[}{\begin{equation}}

\renewcommand*{\]}{\end{equation}}

\def\PRA{{Phys.~Rev.~A} }

\def\JPB{{J.~Phys.~B} }
\def\PRL{{Phys.~Rev.~Lett.} }

\newcommand{\myscaleboxa}[1]{\scalebox{0.65}[0.65]{#1}}
\newcommand{\myscaleboxb}[1]{\scalebox{0.7}[0.7]{#1}}
\newcommand{\myscaleboxc}[1]{\scalebox{1.4}[1.4]{#1}}

\newcommand{\myscaleboxe}[1]{\scalebox{0.75}[0.75]{#1}}
\newcommand{\myscaleboxf}[1]{\scalebox{0.9}[0.9]{#1}}

\usepackage{epsfig}

\usepackage{hyperref}

\begin{document}

\title{Generation of isolated attosecond pulses in the far field by spatial filtering with an intense few-cycle mid-infrared laser}

\author{Cheng Jin}
\affiliation{J. R. Macdonald Laboratory, Physics Department, Kansas
State University, Manhattan, Kansas 66506-2604, USA}

\author{Anh-Thu Le}
\affiliation{J. R. Macdonald Laboratory, Physics Department, Kansas State University, Manhattan, Kansas 66506-2604, USA}

\author{Carlos A. Trallero-Herrero}
\affiliation{J. R. Macdonald Laboratory, Physics Department, Kansas State University, Manhattan, Kansas 66506-2604, USA}

\author{C. D. Lin}
\affiliation{J. R. Macdonald Laboratory, Physics Department, Kansas State University, Manhattan, Kansas 66506-2604, USA}

\date{\today}

\begin{abstract}
We report theoretical calculations of high-order harmonic generation
(HHG) of Xe with the inclusion of multi-electron effects and
macroscopic propagation of the fundamental and harmonic fields in an
ionizing medium. By using the time-frequency analysis we show that
the reshaping of the fundamental laser field is responsible for the
continuum structure in the HHG spectra. We further suggest a method
for obtaining an isolated attosecond pulse (IAP) by using a filter
centered on axis to select the harmonics in the far field with
different divergence. We also discuss the carrier-envelope-phase
dependence of an IAP and the possibility to optimize the yield of
the IAP. With the intense few-cycle mid-infrared lasers, this offers
a possible method for generating isolated attosecond pulses.

\end{abstract}

\pacs{42.65.Ky,42.65.Re,42.65.Jx,31.70.Hq,32.80.Aa}

\maketitle

\section{Introduction}

High-order harmonic generation (HHG), one of the most interesting
nonlinear phenomena occurring when atoms or molecules are exposed to
an intense infrared laser field, has been widely used for the
production of attosecond pulses in the extreme ultraviolet (XUV)
\cite{krausz-rmp-2009,nsoli-pqe-2009,dimauro-rpp-2004,Mette-jpb,pop-natph-2010}.
Due to its great potential for probing ultrafast electronic
processes, different methods have been used to generate isolated
attosecond pulses (IAPs). Using carrier-envelope-phase (CEP)
stabilized few-cycle laser pulses, an IAP as short as 80 attoseconds
has been generated by synthesizing harmonics beyond the cutoff
\cite{christov-prl-1997,science-80as-2008}. Starting with an
elliptically polarized light, the polarization gating technique in
which HHG emission from the central cycle of the pulse is selected,
has also produced an isolated 130-as pulse
\cite{sansone-sci-2006,mashiko-prl-2008}. In a tight focusing
geometry, using the so-called spatiotemporal gating, an IAP can be
generated since different phase-matching conditions can be achieved
for different ranges of harmonics
\cite{haworth-NatPhys-2007,pfeifer-oe-2007}. Other alternative
methods of IAP generation have been reported, including confining
harmonics generated in a narrow temporal window in the leading edge
of a laser pulse, where a 210-as IAP has been reported
\cite{thomann-oe-2009}. The IAP can also be generated by optimizing
the pressure and length of the gas cell
\cite{liu-ol-2010,liu-pra-2009,zheng-apl-2009} or using a spatial
filter in the far field
\cite{krausz-nature-2001,mette-ol-2006,mette-pra-2006}. Indeed there
is a plethora of techniques for the production of the IAP, with the
idea that harmonics be generated from half an optical cycle only in
a few- or multi-cycle infrared laser pulse.

Since the harmonic field generated by all atoms or molecules within
the laser focus co-propagates with the fundamental laser field in
the medium, as well as possible further propagation in the free
space depending on the experimental setup, the understanding of the
observed HHG consists of two parts: first, the HHG emission from
individual atoms (or molecules) through the laser-induced dipole;
second, the propagation of the fundamental and the harmonic fields
in the medium and free space. There are two fundamental equations to
be solved: time-dependent Schr\"{o}dinger equation (TDSE) for the
single-atom (or single-molecule) response, and Maxwell's wave
equation for the propagation process. In practice, in the first
step, the solution of TDSE is normally not employed due to
computational difficulties. Instead, strong-field approximation
(SFA) (or Lewenstein model) \cite{Lewen-pra-1994} is used. However,
it is well-known that SFA fails to reproduce the observed HHG
spectra and thus the reliability of the predicted spectra after
propagation in the medium is questionable. In recent years, a
quantitative rescattering (QRS) theory has been proposed by our
group \cite{lin-jpb-10,toru-2008,at-pra-2009}, which has been
corroborated subsequently by others
\cite{Starace-jpb-2009,Starace-prl-2009,cerkic09,tolstikhin10}. The
single-atom (or single-molecule) response obtained from QRS has been
shown to be as accurate as the one from TDSE, but the calculation is
as easy as SFA. Taking advantage of this theoretical success,
recently we have incorporated the QRS theory into the macroscopic
propagation of harmonic fields in the medium. Until now, we have
shown that QRS-based macroscopic harmonic spectra: (i) agreed well
with TDSE-based HHG spectra for Ar \cite{jin-pra-2009}; (ii)
compared well with experimental HHG spectra reported for Ar,
N$_{2}$, and CO$_{2}$ \cite{jin-jpb-2011,jin-pra-2011,jin-co2} when
the experimental conditions are well specified. It has been
demonstrated that this approach can be used to calculate macroscopic
HHG spectra by polyatomic molecules \cite{zhao-pra-2011} even though
the predictions have not been tested against experiments yet.
Outside of our group, the QRS theory has been applied to study HHG
by the two-color fields in which the propagation effect is included
\cite{midorikawa-prl-2010}. These applications of the QRS theory are
focused on the HHG spectra. In this paper, we focus on the analysis
of attosecond pulse generation which inevitably tests the phases of
the harmonics obtained in our simulation.

Recently, Xe has become a favorite candidate for generating an
intense IAP \cite{nisoli-natph-2010}, studying phase-matching
effects in the generation of high-energy photons \cite{tosa-2011},
and probing the multi-electron dynamics with high-harmonic
spectroscopy \cite{shiner-np-2011}. Ferrari {\it et al.}
\cite{nisoli-natph-2010} reported the generation of a high-energy
160-as IAP using low-order harmonics of Xe from a CEP-stabilized
laser. They used very high laser intensity and very dilute gas so
that the fundamental field was not severely distorted, but the
ground state of atom was depleted very quickly in the leading edge
of the laser pulse. Only low-order harmonics emitted within one half
cycle were used to obtain an IAP. Shiner {\it et al.}
\cite{shiner-np-2011} used a 1.8-$\mu$m laser with a duration of
less than two optical cycles to obtain the HHG spectra of Xe up to
the photon energy of 160 eV. They have shown that HHG spectra
exhibited strong enhancement above about 90 eV. This enhancement is
well-known in photoionization (PI) of Xe due to the presence of a
strong shape resonance from the 4d shell which, through the channel
coupling, modifies the partial PI cross section of the 5p shell of
Xe -- a feature attributed to many-electron effects. According to
QRS, such enhancement is anticipated since partial
photorecombination (PR) cross section (related to photoionization)
enters directly in the laser-induced dipole. To simulate HHG spectra
at high-photon energies, multi-electron effects on the laser-induced
dipoles thus have to be included. Using such dipoles in the QRS, we
simulate the HHG spectra of Xe generated by 1.8-$\mu$m lasers by
including the macroscopic propagation effects.

In this paper, as a check, we simulate the HHG spectra reported in
Shiner {\it et al.} \cite{shiner-np-2011}. Mostly we aim at
reproducing the HHG spectra of Xe observed experimentally in Ref.
\cite{carlos-2011}, which show nearly continuous photon energy
distributions (to be called continuum structure) at high laser
intensities. Such continuum spectra have also been observed in
molecules, like NO \cite{carlos-2011}. From our simulation, we wish
to demonstrate whether IAPs are generated by these harmonics. For
this, we demonstrate how to select different ranges of harmonics to
synthesize an IAP by using a spatial filter in the far field. This
approach is different from that in Ferrari {\it et al.}
\cite{nisoli-natph-2010}, but similar to the analysis in Gaarde {\it
et al.} \cite{mette-ol-2006}. The rest of this paper is arranged as
follows. In Sec. II, we briefly summarize the propagation equations,
wavelet theory for the time-frequency analysis, formulas for the
attosecond pulse generation, and QRS theory including multi-electron
effects. In Sec. III, the theoretical results are presented and
analyzed for different experimental conditions, for IAPs by
synthesizing harmonic orders from 40 to 80 (H40-H80) and H90-H130.
We also compare attosecond pulses calculated using the QRS and the
SFA. An analysis of CEP dependence of the generated IAP presented at
the end of this section concludes that it is still possible to
obtain an IAP even for lasers where the CEP is not stabilized. A
short summary in Sec. IV concludes this paper.

\section{Theoretical methods}

\subsection{Propagation equations of fundamental and harmonic fields}
The propagation of the fundamental laser field and high harmonics in
an ionizing medium has been described in detail in Ref.
\cite{jin-pra-2011}, so we only recall the main equations here. The
evolution of the fundamental field is described by a
three-dimensional (3D) Maxwell's wave equation
\cite{Tosa-pra-2003,Geissler-prl-99}:
\begin{eqnarray}
\label{fund-time}\nabla^{2}E_{1}(r,z,t)&&-\frac{1}{c^{2}}\frac{\partial^{2}E_{1}(r,z,t)}{\partial
t^{2}}= \mu_{0}\frac{\partial J_{\text {abs}}(r,z,t)}{\partial
t}\nonumber\\&&+\frac{\omega_{0}^{2}}{c^{2}}(1-\eta_{\text
{eff}}^{2})E_{1}(r,z,t),
\end{eqnarray}
where $E_{1}(r,z,t)$ is the transverse electric field with central
frequency $\omega_{0}$.
$\nabla^{2}=\nabla_{\bot}^{2}+\partial^{2}/\partial z^{2}$ in
cylindrical coordinates, where $z$ is the axial propagation
direction. The effective refractive index is
\begin{eqnarray}
\label{eff}\eta_{\text
{eff}}(r,z,t)=\eta_{0}(r,z,t)+\eta_{2}I(r,z,t)-\frac{\omega_{\text
p}^{2}(r,z,t)}{2\omega_{0}^{2}}.
\end{eqnarray}
The linear term $\eta_{0}=1+\delta_{1}-i\beta_{1}$ accounts for
refraction ($\delta_{1}$) and absorption ($\beta_{1}$) by the
neutral atoms, the second term describes the optical Kerr
nonlinearity which depends on the instantaneous laser intensity
$I(t)$, and the third term contains the plasma frequency
$\omega_{\text p}=[e^{2}n_{\text e}(t)/(\varepsilon_{0}m_{\text
e})]^{1/2}$, where $m_{\text e}$ and $e$ are the mass and charge of
an electron, respectively, and $n_{\text e}(t)$ is the density of
free electrons. The absorption term $J_{\text {abs}}(t)$ due to the
ionization of the medium is given by \cite{Mette-jpb,Rae-pra-1992}
\begin{eqnarray}
\label{abs}J_{\text {abs}}(t)=\frac{\gamma(t)n_{\text e}(t)I_{\text
p}E_{1}(t)}{|E_{1}(t)|^{2}},
\end{eqnarray}
where $\gamma(t)$ is the ionization rate, and $I_{\text p}$ is the
ionization potential. Ionization rates involved in Eq. (\ref{abs})
and in free electron density $n_{\text e}(t)$ are calculated using
the improved Ammosov-Delone-Krainov (ADK) theory \cite{tong-jpb}.
The fundamental laser field is assumed to be Gaussian both in space
and time at the entrance of the medium, and the gas pressure is
constant within the medium.

The 3D propagation equation of the harmonic field is
\cite{Priori-pra-2000,Mette-jpb,tosa-pra-2005}
\begin{eqnarray}
\label{harm-time}\nabla^{2}E_{\text
h}(r,z,t)-\frac{1}{c^{2}}\frac{\partial^{2}E_{\text
h}(r,z,t)}{\partial
t^{2}}=\mu_{0}\frac{\partial^{2}P(r,z,t)}{\partial t^{2}},
\end{eqnarray}
where $P(r,z,t)$ is the polarization depending on the applied
fundamental field $E_{1}(r,z,t)$. Here the free-electron dispersion
is neglected because the frequencies of high harmonics are much
higher than the plasma frequency. In general, the polarization
$P(r,z,t)$ is separated into linear and nonlinear components, and
the linear susceptibility $\chi^{(1)}(\omega)$ includes both linear
dispersion and absorption effects of the harmonics \cite{Henke}. The
nonlinear polarization term $P_{\text {nl}}(r,z,t)$ can be expressed
as
\begin{eqnarray}
\label{pola}P_{\text {nl}}(r,z,t)=[n_{0}-n_{\text
e}(r,z,t)]D(r,z,t),
\end{eqnarray}
where $n_{0}-n_{\text e}(r,z,t)$ gives the density of the remaining
neutral atoms, and $D(r,z,t)$ is the single-atom induced dipole
moment. Note that Eqs. (\ref{fund-time}) and (\ref{harm-time}) are
solved using the Crank-Nicholson routine in the frequency domain.

Once the harmonic field at the exit face (near field) of the medium
is computed, the harmonics propagating in free space in the far
field can be obtained from near-field harmonics through a Hankel
transformation \cite{far-field,tosa-2009,chipperfield-2009}.

\subsection{Wavelet analysis of attosecond pulses}
A time-frequency representation (TFR) (or spectrogram) of the
harmonic field $E_{\text h}(t)$ is a simultaneous representation of
the temporal and spectral characteristics of the harmonics. We
perform the time-frequency analysis in terms of the wavelet
transform of the harmonic field
\cite{tong-pra-2000,mette-pra-1998,mette-oe-2001,yakovlev-prl-2003}:
\begin{eqnarray}
A(t,\omega)=\int E_{\text h}(t')w_{t,\omega}(t')dt',
\end{eqnarray}
with the wavelet kernel
$w_{t,\omega}(t')=\sqrt{\omega}W[\omega(t'-t)]$. We choose the
Morlet wavelet \cite{tong-pra-2000}:
\begin{eqnarray}
W(x)=(1/\sqrt{\tau})e^{ix}e^{-x^{2}/2\tau^{2}}.
\end{eqnarray}
The width of the window function in the wavelet transform varies as
the frequency changes, but the number of oscillations (proportional
to $\tau$) within the window is held constant. The dependence of
$A(t,\omega)$ on the parameter $\tau$ has been tested. The absolute
value of $A(t,\omega)$ depends on $\tau$, but the general temporal
pattern does not change much. In this paper, we choose $\tau=15$ to
perform the wavelet transform.

Harmonics emitted at the exit plane (near field) of the medium act
as a source for the far-field harmonics. In order to avoid the
complexity of the harmonic spatial distribution in the near field
(see Fig. 4 in Ref. \cite{jin-pra-2011}), we calculate $A(t,\omega)$
for each radial point in the near field and then integrate over the
radial coordinate \cite{mette-oe-2001}:
\begin{eqnarray}
\label{near-tf}|A_{\text {near}}(t,\omega)|^{2}=\int_{0}^{\infty}
2\pi r dr |\int E_{\text h}(r,t')w_{t,\omega}(t')dt'|^{2}.
\end{eqnarray}
To demonstrate the divergence of harmonics, we preform the TFR for
each radial point in the far field.

The spectral filter used to select a range of harmonics ($\omega_1$
- $\omega_2$) could affect the generation of attosecond pulse trains
(APTs) or IAPs. Theoretically we can obtain the total intensity of
an APT or an IAP in the near field as following
\cite{antoine-prl-1996}:
\begin{eqnarray}
\label{near-int}I_{\text {near}}(t)=\int_{0}^{\infty} 2 \pi r
dr|\int_{\omega_{1}}^{\omega_{2}}E_{\text h}(r,\omega)e^{i\omega
t}d\omega|^{2}.
\end{eqnarray}
In the far field, a spatial filter is used to select the harmonics
in a prescribed area. In this paper, we assume that the filter is
circular with a radius $r_{0}$, and is perpendicular to the
propagation direction of harmonics. The intensity of an APT or an
IAP in the far field is
\begin{eqnarray}
\label{far-int}I_{\text {far}}(t)=\int_{0}^{r_{0}} 2 \pi r
dr|\int_{\omega_{1}}^{\omega_{2}}E_{\text h}^{\text
f}(r,\omega)e^{i\omega t}d\omega|^{2}.
\end{eqnarray}

\subsection{Photorecombination (PR) dipole moment of Xe in the QRS theory}
The single-atom induced dipole moment $D(t)$ in Eq. (\ref{pola}) is
obtained by the QRS theory. It can be expressed in energy (or
frequency) domain as following \cite{at-pra-2009,at-pra-2008}:
\begin{eqnarray}
\label{single-qrs}D(\omega)=W(\omega)d(\omega),
\end{eqnarray}
where $d(\omega)$ is the PR transition dipole moment and $W(\omega)$
is the microscopic wave packet. In QRS theory, $W(\omega)$ is
determined by the laser field and can be accurately calculated from
SFA, and $d(\omega)$ is the transition dipole between the initial
and final states of PR (or PI). When the multi-electron effect is
not important, the transition dipole can be calculated using the
single-active electron (SAE) approximation. However, the transition
dipole is easily generalized to include many-electron effects, as
routinely done in PI theory of atoms and molecules. Thus to include
many-electron effects in $d(\omega)$, multi-channel calculations
such as many-body perturbation theory, close-coupling method,
R-matrix method, random-phase approximation, and many others can all
be employed for such purpose. Since PI of Xe has been well studied,
we obtain $d(\omega)$ semi-empirically. The major many-body effect
for PI of Xe from 5p shell occurs at photon energy where 4d shell is
open. Thus below about 60 eV, the transition dipole from 5p can be
obtained from a single-electron model. This gives the magnitude and
phase of the transition dipole. At higher energies, effects from the
4d shell on the transition dipole of 5p becomes important since PI
cross section of Xe from 4d has a large and broad shape resonance
around 100 eV. The intershell coupling will enhance the $d(\omega)$
for 5p near and above 90 eV. Such enhancement has been calculated in
Kutzner {\it et al.} \cite{kutzner-pra-1989} using the relativistic
random-phase approximation (RRPA). In our calculation, the phase of
$d(\omega)$ is taken from the 5p shell under the SAE approximation
\cite{jin-pra-2011} while the magnitude is taken from Ref.
\cite{kutzner-pra-1989}. This approximation does not change the
temporal structure of attosecond pulses (shown later) since the
phase of $D(\omega)$ is dominated by the phase of the wave packet
$W(\omega)$. We comment that in QRS the induced dipole is given in
the energy domain, thus the calculation is similar to the time
independent theory used in PI which has been well-established in the
last 30 years.

\section{Results and discussion}
\subsection{Macroscopic HHG spectra of Xe at different laser intensities}
HHG spectra of Xe extended to the photon energy of over one hundred
electron volts using 1.8-$\mu m$ lasers with the pulse duration of
few optical cycles have been reported recently
\cite{shiner-np-2011,carlos-2011}.

\begin{figure}
\mbox{\rotatebox{270}{\myscaleboxb{
\includegraphics[trim=0mm 8mm -2mm -4mm, clip]{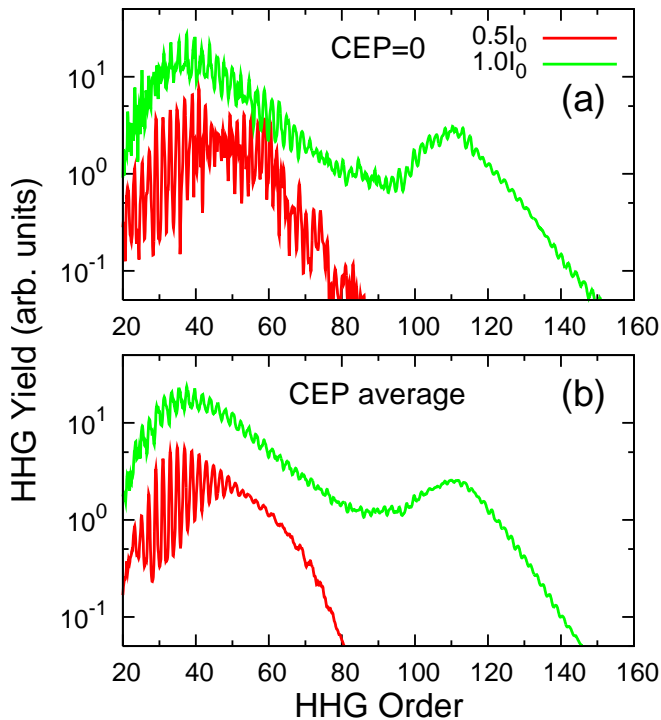}}}}
\caption{(Color online) Macroscopic HHG spectra of Xe in an 1825-nm
laser, for (a) CEP=0 and (b) CEP averaged. Laser intensities are
indicated in units of I$_{0}$=10$^{14}$ W/cm$^{2}$. See text for
additional laser parameters and the experimental arrangement.
\label{Fig1}}
\end{figure}

In Fig.~\ref{Fig1}, we show the calculated HHG spectra of Xe exposed
to a 14-fs (FWHM), 1825-nm laser. The laser beam waist is 100
$\mu$m. A 1-mm-long gas jet with the pressure of 30 Torr is placed
at the laser focus. The harmonics are detected after a slit with a
width of 190 $\mu$m and placed 455 mm behind the focus. These
parameters are chosen to be close to those in the experiment of
Trallero-Herrero {\it et al.} \cite{carlos-2011}. For the present
purpose we analyze HHG spectra obtained from our theoretical
simulations at two laser peak intensities 0.5$\times$10$^{14}$
W/cm$^{2}$ and 1.0$\times$10$^{14}$ W/cm$^{2}$, which are below and
above the critical intensity for Xe at $\sim$ 0.87$\times$10$^{14}$
W/cm$^{2}$ \cite{tong-jpb}, respectively. Here the critical
intensity is defined with respect to the static electric field where
an electron can escape over the top of the field-induced potential
barrier classically.

We show the macroscopic HHG spectra for CEP=0 in Fig.~\ref{Fig1}(a).
The two laser intensities present different characteristics of
harmonics. For the low intensity, the harmonics are very sharp,
i.e., the valley between the neighboring odd harmonics is very deep.
At high intensity, the valley is very shallow, i.e., the spectrum
shows a continuum structure. Furthermore, the harmonics are not
exactly at odd orders due to the blue shift of the fundamental
field. Note that the spectrum rises above about H90 is due to the
intershell or many-electron effects discussed in Sec. II C. Since a
few-cycle laser pulse is applied, the HHG spectra have a strong CEP
dependence. In Fig.~\ref{Fig1}(b), we show the CEP averaged HHG
spectra. The main characteristics of harmonics remain the same
except that the harmonic spectra are much smoother. The CEP is fixed
at zero in the following sections unless otherwise stated.

\subsection{Spatiotemporal evolution of the fundamental laser field}
\begin{figure*}
\mbox{\rotatebox{270}{\myscaleboxb{
\includegraphics[trim=-5mm 8mm -1mm 5mm, clip]{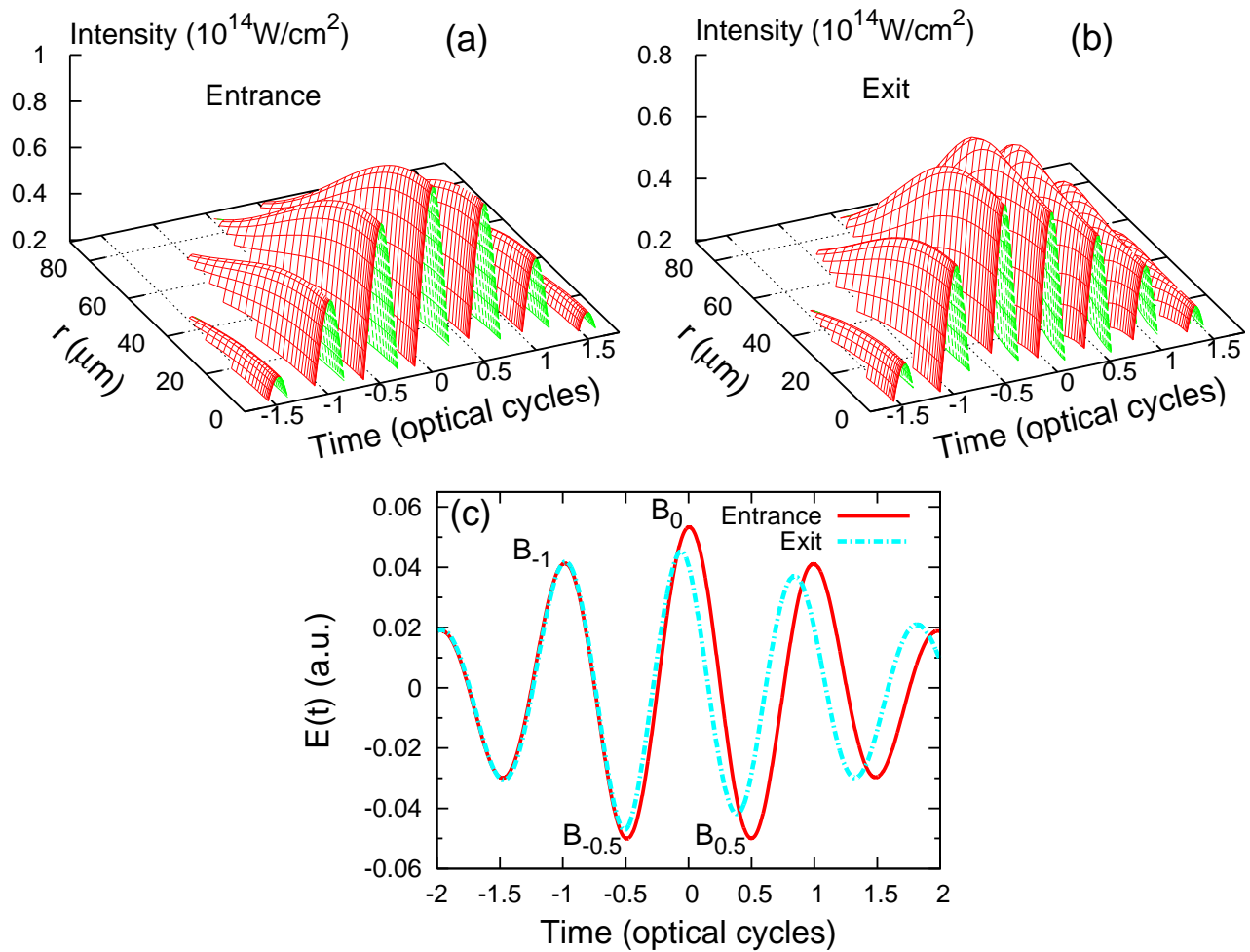}}}}
\caption{(Color online) Spatiotemporal intensity profile of the
laser pulse at (a) the entrance and (b) the exit of Xe gas jet.
Laser intensity at the focus is 1.0$\times$10$^{14}$ W/cm$^{2}$
(assumed in the vacuum) and CEP=0. (c) Evolution of the on-axis
electric field at the entrance (solid line) and the exit (dot-dashed
line). The laser field becomes chirped during the propagation. For
sub-cycle dynamics analysis, we use the label B$_{t}$, with t=-1,
-0.5, 0, and 0.5 (in units of optical cycles) to indicate the
approximate half-cycle where the electron is born. Note that t is
defined within the half cycle only. \label{Fig2}}
\end{figure*}

To understand the different spectral features in Fig.~\ref{Fig1}, we
inspect the fundamental field in the ionizing medium. The
spatiotemporal intensity profile and on-axis electric fields of the
laser pulse at the entrance and the exit of the gas jet are shown in
Fig.~\ref{Fig2}. The laser peak intensity is 1.0$\times$10$^{14}$
W/cm$^{2}$, which would give an ionization probability of
$\sim$35$\%$ at the end of laser pulse for Xe according to an
empirical ADK formula in barrier-suppression regime \cite{tong-jpb}.
While the electric field at the entrance has a good Gaussian shape
both in time and space, it is strongly reshaped during the
propagation in the ionizing medium. At the exit it shows positive
chirp in time (blue shift in frequency) [see Fig.~\ref{Fig2}(c)] and
defocusing in space [see Fig.~\ref{Fig2}(b)]. We have also checked
the fundamental field with laser peak intensity of
0.5$\times$10$^{14}$ W/cm$^{2}$. It always maintains Gaussian
spatial distribution and there is no blue shift because the
ionization probability is very low. The reshaping of the fundamental
field at high intensity is responsible for the continuum structure
in the HHG spectra in Fig.~\ref{Fig1}. Note that similar results
have been obtained by Gaarde {\it et al.} \cite{mette-pra-2006}
using a 750-nm laser interacting with Ne gas.

\subsection{Time-frequency analysis of harmonics in the near and far fields}
\begin{figure*}
\mbox{\rotatebox{270}{\myscaleboxe{
\includegraphics[trim=-14mm 9mm -4mm 14mm, clip]{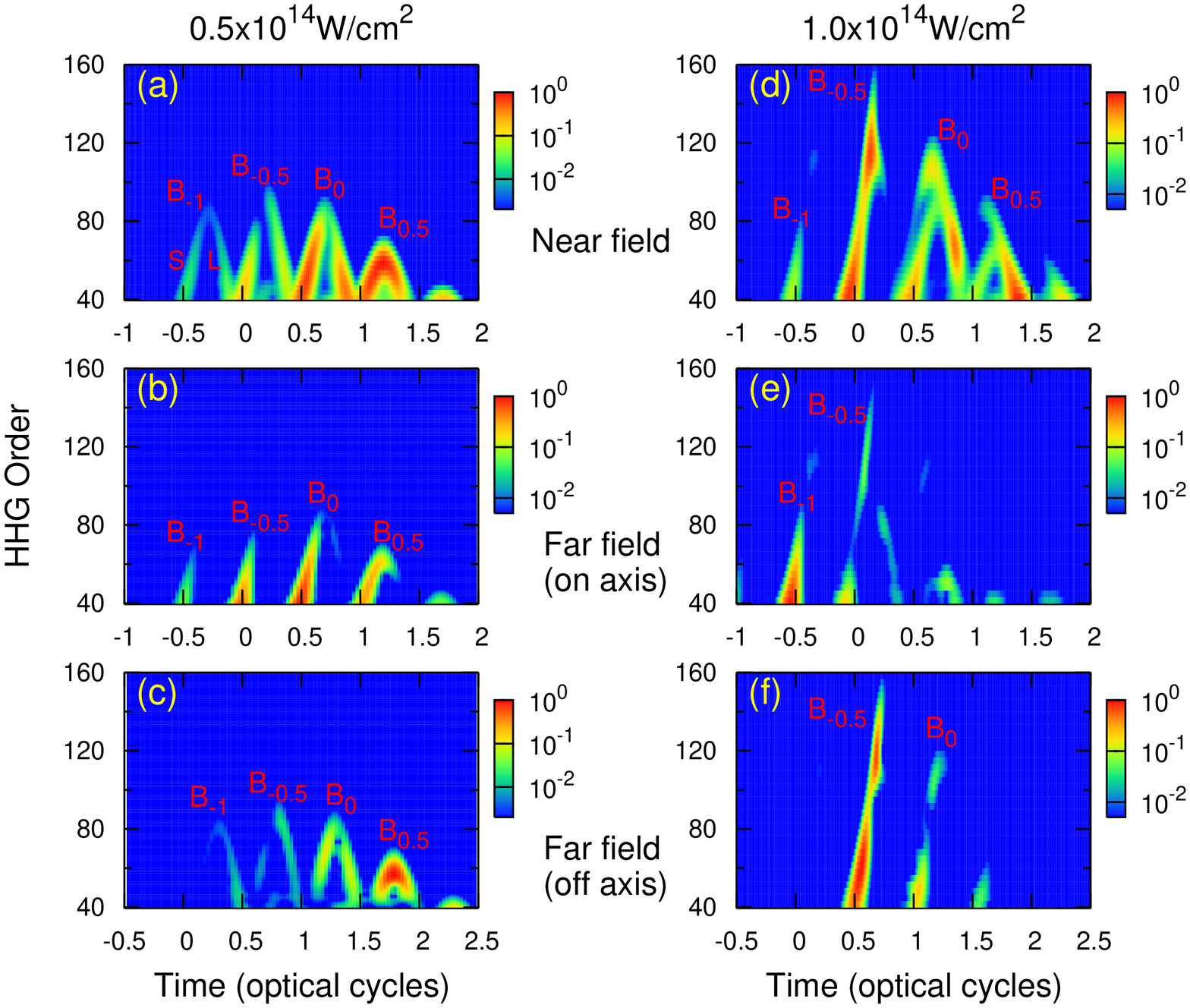}}}}
\caption{(Color online) Top row: Time-frequency representation (TFR)
of harmonics in the near field. Middle row:  TFR for on-axis (r=0
mm, divergence: 0 mrad) harmonics in the far field. Bottom row: TFR
for off-axis (r=1 mm, divergence: 2.2 mrad) harmonics in the far
field. Far-field position is at z=455 mm, and laser intensity
(CEP=0) along each column is indicated. Electrons are released at
each half cycle, labeled by B$_{t}$, with t=-1, -0.5, 0, and 0.5 as
in Fig.~\ref{Fig2}. For each B$_{t}$, electrons can follow a
``short" (S) or ``long" (L) trajectory to recombine with the ion to
emit harmonics. For each harmonic, the emission time can be read
from the time axis. For each B$_{t}$, the emission time for each
off-axis harmonic is delayed with respect to the corresponding
on-axis harmonic, e.g., compare (b) vs (c), and (e) vs (f). All the
TFRs have been normalized. \label{Fig3}}
\end{figure*}

Harmonic generation is a temporal coherent process and can be better
understood if we study it in terms of its emission time. In this
subsection we examine the time-frequency representation (TFR) of
harmonics in the near and far fields for the low and high laser
intensities.

For each harmonic order $q$, it is known that the phase can be
expressed as \cite{mette-pra-1999}:
\begin{eqnarray}
\Phi_{i}^{q}(r,z,t)=-\alpha_{i}^{q}I(r,z,t),
\end{eqnarray}
where $I(r,z,t)$ is the spatiotemporal intensity of the fundamental
laser field. The proportional constant $\alpha_{i=S,~L}$ depends on
``short" (S) or ``long" (L) trajectories. The phase can also be
expressed in terms of the ponderomotive energy $U_{p}$ and the
electron excursion time $\tau_{i}^{q}$: $\Phi_{i}^{q} \approx
-\beta_{i} U_{p} \tau_{i}^{q}$ \cite{zair-prl-2008}, where the
coefficient $\beta_{i}$ for the ``short" trajectory is much smaller
than for the ``long" trajectory. The electron excursion times for
the two trajectories are $\tau_{S}^{q} \approx T/2$ and
$\tau_{L}^{q} \approx T$ ($T$ is the laser period)
\cite{lewen-pra-1995}. It shows that the phase grows with the cubic
power of the wavelength. The curvature of the phase front caused by
the radial variation $\partial \Phi_{i}^{q}(r)/\partial r$ makes the
harmonic beam divergent. The divergence of ``short"- or
``long"-trajectory harmonic is determined by either $\Delta
\alpha_{i}^{q}$ or $\Delta I(r)$.

\subsubsection{Harmonics in the near field}
The TFR, $|A_{\text {near}}(t,\omega)|^{2}$, calculated from
Eq.~(\ref{near-tf}), are shown in Figs.~\ref{Fig3}(a) and (d) for
harmonics above H40 at two laser intensities, collected at the exit
face of the gas jet (near field). In Fig.~\ref{Fig3}(a), the symbols
S and L are used to indicate the first (earliest) group of harmonics
generated. Here S (L) stands for ``short" (``long")-trajectory
harmonics that have positive (negative) chirp. These harmonics are
from electrons born at t=-1 (in units of optical cycles), i.e.,
B$_{-1}$ to indicate born time at t=-1,  in the leading edge of the
pulse [see Fig.~\ref{Fig2}(c)]. In the following, the electron born
time t (in units of optical cycles) is indicated by B$_{t}$ in the
figure, while the harmonic emission time is read off from the
horizontal axis of the figure, one for the ``short", and the other
for the ``long" trajectory. In this paper the time is always defined
in moving coordinate frame \cite{jin-pra-2011}. At the low intensity
in Fig.~\ref{Fig3}(a), we can see that both S and L contribute to
harmonics generated from electrons born at t=-1, -0.5, 0, and 0.5.
In other words, harmonics are generated by electrons born over four
half cycles. Note that Tate {\it et al.} \cite{kate-prl-2007} have
shown that harmonics generated by mid-infrared lasers had large
contributions from electron trajectories even longer than the
``long" trajectories in single-atom response, which has also been
confirmed in our calculation (not shown). But these trajectories are
all eliminated during the propagation in the medium since their
phases are very large. For low intensity,   the propagation in the
medium cannot eliminate contributions from ``long" trajectories.

The same TFR analysis for the high intensity is shown in
Fig.~\ref{Fig3}(d). Higher harmonic cutoff from each burst is easily
seen since the intensity is twice higher. Comparing to
Fig.~\ref{Fig3}(a), there are no contributions to the harmonics from
the ``long" trajectories for electrons born at t=-1 and -0.5, i.e.,
from the leading edge of the pulse. Since the laser intensity is
twice higher, the phase of each harmonic is also twice higher (also
see Fig. 17 in Ref. \cite{sansone-pra-2004} and Fig. 1(A) in Ref.
\cite{mairesse-sci-2003}), thus resulting in cancelation of
contributions from the ``long" trajectories. For electrons born at
the falling edge of the pulse, due to the blue shift (thus shorter
wavelength) and reshaping (thus lower intensity) the phases of
harmonics due to the ``long" trajectories are smaller and they can
survive after propagation in the medium, for example, for electrons
born at t=0, and 0.5, see Fig.~\ref{Fig3}(d).

\subsubsection{On-axis harmonics in the far field}

In Fig.~\ref{Fig3}(b), the TFR is shown for r=0 mm in the far field
(455 mm after the laser focus). At low intensity, the emission from
``short" trajectories born at different times have the similar small
divergence, and after propagation in free space they all survive
along the axis in the far field. Interference between
``short"-trajectory harmonics from each half cycle leads to
enhancement in odd harmonics and suppression in even harmonics (see
Fig. 18 in Ref. \cite{knight-1997}), and resulting in a big contrast
between an odd harmonic and neighboring harmonics shown in
Fig.~\ref{Fig1}(a) for the spectra obtained with a slit. At high
intensity, only harmonics from the ``short"-trajectory electrons
born at t=-1 survive (the next one at t=-0.5 is much weaker), see
Fig.~\ref{Fig3}(e). This would result in a nearly continuum spectra
and a potential for generating an isolated attosecond pulse.

\subsubsection{Off-axis harmonics in the far field}
In Figs.~\ref{Fig3}(c) and (f), the TFR is shown for r=1 mm
(divergence: 2.2 mrad) in the far field. Each off-axis burst has an
obvious time delay with respect to the on-axis burst because it
travels a longer distance in free space. At low intensity, harmonics
from ``long" trajectories appear on each burst since they have large
divergence [see Fig.~\ref{Fig3}(c)]. At high intensity, the ``short"
trajectories contribute to bursts B$_{-0.5}$ and B$_{0}$ [see
Fig.~\ref{Fig3}(f)]. They appear to come from the pulse reshaping,
see Fig.~\ref{Fig2}(b) showing laser peak intensity shifting to
region away from the propagation axis. They experience larger
$\Delta I(r)$ with respect to ``short"-trajectory electrons born at
B$_{-1}$ at the leading edge. Fig.~\ref{Fig3}(f) shows that a
continuum spectra from a ``short" trajectory is generated for
electrons born at t=-0.5.

Note that attochirp (emission time varying with harmonic order)
\cite{mairesse-sci-2003,mairesse-prl-2004} of ``short"- or
``long"-trajectory harmonics exists even after propagation. They may
be compensated using a ``plasma compressor" \cite{mairesse-sci-2003}
because free electrons induce  a negative group velocity dispersion,
or by thin filters with linear negative group velocity dispersion
\cite{compen-prl-2005}. But attochirp is inversely proportional to
laser wavelength \cite{doumy-prl-2009}. This implies that one can
select a broad range of harmonics to synthesize a short attosecond
pulse using an 1825-nm laser (will be shown next). The harmonic
emission of ``short" trajectory in the far field in
Figs.~\ref{Fig3}(e) and (f) varies with time or radial distance.
This provides the possibilities to generate IAPs using the different
ranges of harmonics on or off axis. We will only show the spectral
and spatial filters applied on axis in the far field in the
following.

\subsection{Spectral and spatial filtering in the generation of attosecond pulses}
\begin{figure*}
\mbox{\rotatebox{270}{\myscaleboxb{
\includegraphics[trim=7mm 1mm 6mm -3mm, clip]{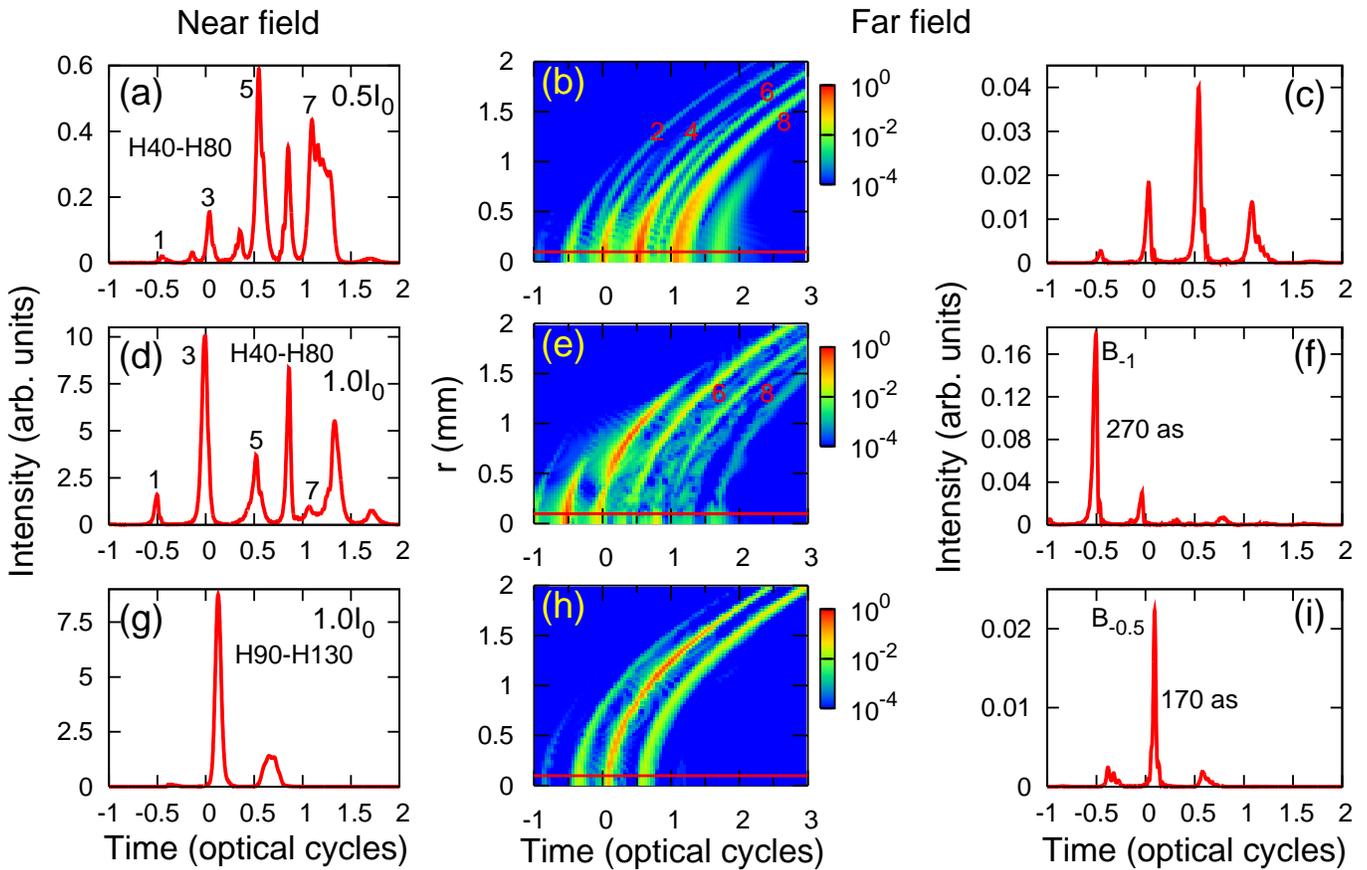}}}}
\caption{(Color online) First column: Intensity (or envelope) of
attosecond pulses in the near field, synthesized from the harmonics
and the laser intensity  shown in each frame. Laser intensities are
given in units of I$_{0}$=10$^{14}$ W/cm$^{2}$. In (a) and (d), odd
bursts (``short" trajectories) are labeled. Even bursts due to
``long" trajectories are not labeled for brevity. Middle column:
Spatial distribution (normalized) of attosecond pulses in the far
field (z=455 mm). Notice that even bursts (``long" trajectories)
have large divergence, or at large r. The odd bursts (not labeled)
have smaller divergence. There is a time delay between off-axis
attosecond pulses compared to on-axis ones. Last column: Intensity
of attosecond pulses in the far field using a spatial filter with a
radius r$_{0}$=100 $\mu$m (shown by the solid line in red in each
middle-column frame). \label{Fig4}}
\end{figure*}

A spectral filter is usually used to synthesize attosecond pulses.
In this section we also study how the attosecond pulses are
manipulated through spatial filtering. Fig.~\ref{Fig4}(a) displays
the intensity profile of an XUV light by synthesizing H40-H80 at the
near field generated by laser intensity of 0.5$\times$10$^{14}$
W/cm$^{2}$. The intensity of the attosecond pulses I$_{near}$(t) is
calculated by using Eq.~(\ref{near-int}). The time-frequency
analysis of these harmonics has been given in Fig.~\ref{Fig3}(a).
Besides attosecond bursts occurring at each half optical cycles,
which can be attributed to harmonics resulting from ``short"
trajectories, we observe other pulses in between which are
attributed to contributions from ``long" trajectories. The main
peaks from the ``short" trajectories are labeled by 1, 3, 5, and 7
in the figure, while those in between (2, 4, 6, and 8 are not
labeled) are from ``long" trajectories. The attosecond pulses thus
generated show a poor periodicity in time, see Fig.~\ref{Fig4}(a).

If the XUV light is synthesized at the far field, in particular, by
introducing a spatial filter, then it may be possible to remove
harmonics resulting from the ``long" trajectories. In
Fig.~\ref{Fig4}(b), the intensity distributions of the synthesized
light in space in the far field are shown. They are obtained from
the near-field harmonics by further propagation in free space. The
peaks 2, 4, 6, and 8 [not shown in Fig.~\ref{Fig4}(a) explicitly]
are attributed to ``long" trajectories. They are indicated in
Fig.~\ref{Fig4}(b) showing that they are distributed far from the
propagation axis. By using a spatial filter (indicated by a solid
line in red, with a radius r$_0$=100 $\mu$m) to select harmonics
generated near the axis only, as shown in Fig.~\ref{Fig4}(c) by
using Eq.~(\ref{far-int}) to calculate I$_ {far}$(t), well-behaved
APTs are then obtained. We comment that the time delay between
off-axis and on-axis harmonics leads to the curved spatial
distribution in Fig.~\ref{Fig4}(b), and it can be understood
mathematically since each harmonic behaves like a Gaussian beam, and
the geometric phase of each harmonic is proportional to r$^{2}$
along the transverse direction (see Fig. 4 in Ref.
\cite{jin-pra-2011}). The traveling distance of off-axis harmonics
can be compensated using a reflecting mirror to refocus the harmonic
beam or by a detector with a curved surface. In principle, this
compensation becomes important to reduce the duration of attosecond
pulses when a spatial filter with a large radius is applied. In this
paper, the radius of the spatial filter is chosen to be small enough
to avoid this curvature effect.

Next we use the same range of harmonics (H40-H80) generated by the
laser intensity of 1.0$\times$10$^{14}$ W/cm$^{2}$ to synthesize
attosecond pulses in the near field. Referring to
Fig.~\ref{Fig3}(b), the ``short" trajectories dominate the harmonic
generation in the leading edge of the laser, while the ``long"
trajectories dominate the harmonic generation in the falling edge.
The synthesized XUV light, shown in Fig.~\ref{Fig4}(d) indeed
reflects this point where the first two peaks occur at multiples of
half optical cycles, while the last four peaks are not. In
Fig.~\ref{Fig4}(e), the spatial distribution of the synthesized XUV
light in the far field indeed supports this description. By using a
spatial filter (indicated by a solid line in red, with a radius
r$_0$=100 $\mu$m) to select only ``short" trajectories, as shown in
Fig.~\ref{Fig4}(f), a nice IAP with a duration of 270 as is
obtained, accompanied by a weak sub-pulse with a much weaker
intensity. This demonstrates the generation of IAPs using spatial
filtering. A similar mechanism of IAP generation has been proposed
by Strelkov {\it et al.} \cite{strelkov-2009,strelkov-njp-2008}
using the harmonics in the plateau region generated by the Ar gas
with very high pressure.

The TFR in Fig.~\ref{Fig3}(e) shows considerable on-axis emission
above H80 at burst B$_{-0.5}$. We use H90-H130 to generate
attosecond pulses in the near field in Fig.~\ref{Fig4}(g). Both
bursts have considerable contributions from ``short" trajectories.
In the far field [see Fig.~\ref{Fig4}(h)], they show different
divergences as discussed before. Finally, we obtain an IAP with a
duration of about 170 as in Fig.~\ref{Fig4}(i) with a spatial
filter. The intensity of the IAP is about 1/8 as that in
Fig.~\ref{Fig4}(f) due, not only to the larger divergence of
``short"-trajectory harmonics born at B$_{-0.5}$ than at B$_{-1}$,
but also the lower harmonic intensity of H90-H130 than that of
H40-H80. On the other hand, the duration of the IAP is decreased.
Similar mechanism of IAP generation has been proposed by Gaarde {\it
et al.} \cite{mette-ol-2006,mette-pra-2006} using harmonics in the
cutoff region by a 750-nm laser exposed on Ne gas.

\begin{figure*}
\mbox{\rotatebox{270}{\myscaleboxb{
\includegraphics[trim=-4mm -1.5mm 7mm -5.5mm, clip]{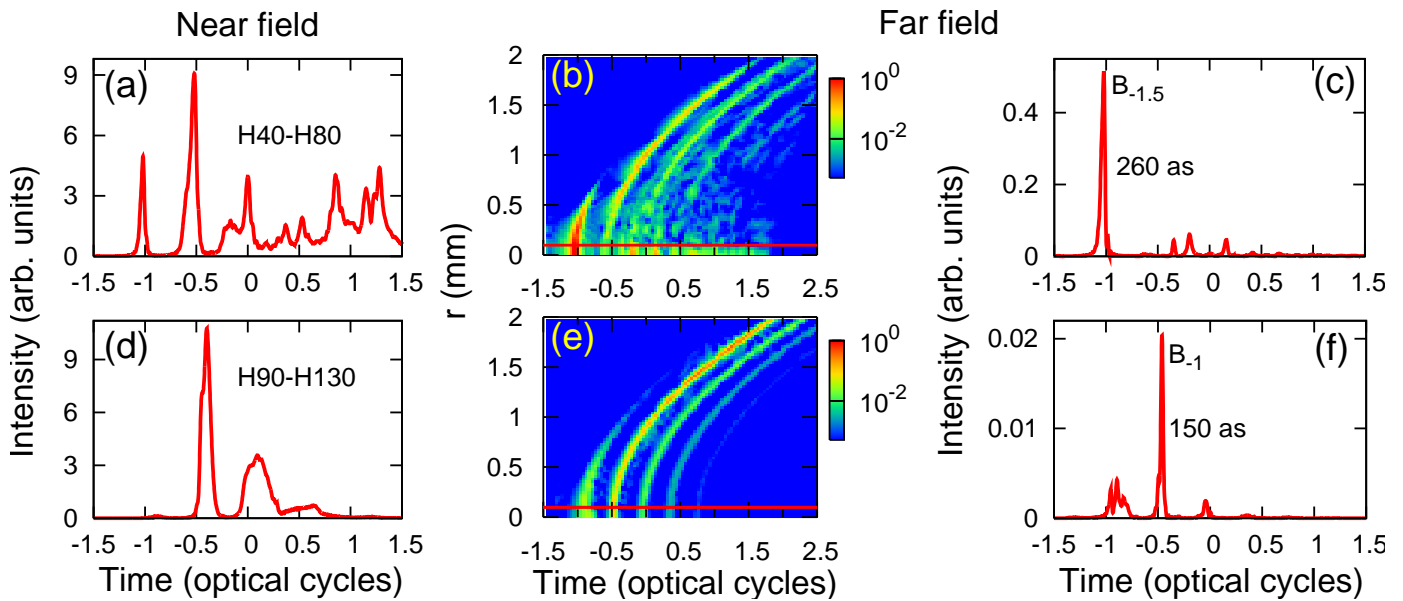}}}}
\caption{(Color online) Top row: Attosecond pulses synthesized from
H40-H80, (a) at the near field, (b) its spatial distribution at the
far field (z=455 mm), and (c) a good isolated attosecond pulse
generated if a filter with a radius r$_{0}$=100 $\mu$m is used to
select only near-axis harmonics (indicated by a solid line in red in
the middle frame). Bottom row: Attosecond pulses synthesized from
H90-H130, (d), (e), and (f) are similar to (a), (b), and (c). The
born times for the bursts that generate the IAP for H40-80 and for
H90-130 are indicated by B$_{t}$ for t=-1.5 and -1, respectively.
Laser intensity is 2.0$\times$10$^{14}$ W/cm$^{2}$ and
CEP=0.\label{Fig5}}
\end{figure*}

We next check how the IAP generation works at higher intensities,
say at 2.0$\times$10$^{14}$ W/cm$^{2}$. Fig.~\ref{Fig5} should be
compared to Figs.~\ref{Fig4}(d)-(i) directly. For the synthesized
H40-H80 pulse, Fig.~\ref{Fig5}(a) shows that the harmonics are
emitted about half an optical cycle earlier than the one at half the
intensity (1.0$\times$10$^{14}$ W/cm$^{2}$). Fig.~\ref{Fig5}(b)
shows that the pulses generated at the falling edge of the laser
pulse have large divergence and thus they tend to come from ``long"
trajectories. In fact, this portion of the pulse does not have good
periodic time dependence. Fig.~\ref{Fig5}(b) also shows that only
the pulse emitted at t=-1 (in units of optical cycles) is near the
axis, thus a filter selecting near axis harmonics results in an IAP,
as illustrated in Fig.~\ref{Fig5}(c). The IAP has a duration of 260
as. For pulses synthesized from H90-H130, Fig.~\ref{Fig5}(d) shows
that there are two bursts emitted at t=-0.5 and 0, and their lateral
profiles in the far field are shown in Fig.~\ref{Fig5}(e). By using
a filter, an attosecond pulse of 150 as can be obtained. In this
case, the IAP intensity does not increase since the fundamental
laser field is much reshaped and the harmonic has a much bigger
divergence in the far field in comparison with 1.0$\times$10$^{14}$
W/cm$^{2}$. It is concluded that ionization gating still works at a
higher laser intensity, and it is more efficient to select bursts
that are born in the leading edge before the laser field starts to
be depleted and blue shifted.

Another question arises is whether the strength of attosecond pulses
can be improved by increasing gas pressure. For ``weak" fields and
low pressure, Shiner {\it et al.} \cite{shiner-prl-2009} have shown
experimentally that the harmonic yield increases quadratically with
the pressure. For ``high" fields addressed here, laser pulse
reshaping is important, we have confirmed theoretically that
increasing the gas pressure while maintaining the same intensity
would not always increase the harmonic yields of Ar \cite{wang}. On
the other hand, the pressure effect on the high harmonics of Xe at
``high" field has been studied in Ref. \cite{carlos-2011}. But its
effect on the attosecond pulse wasn't examined in this paper.

\subsection{Far-field position dependence of isolated attosecond pulses}
\begin{figure*}
\mbox{\rotatebox{270}{\myscaleboxf{
\includegraphics[trim=0mm -1.5mm 2mm 1mm, clip]{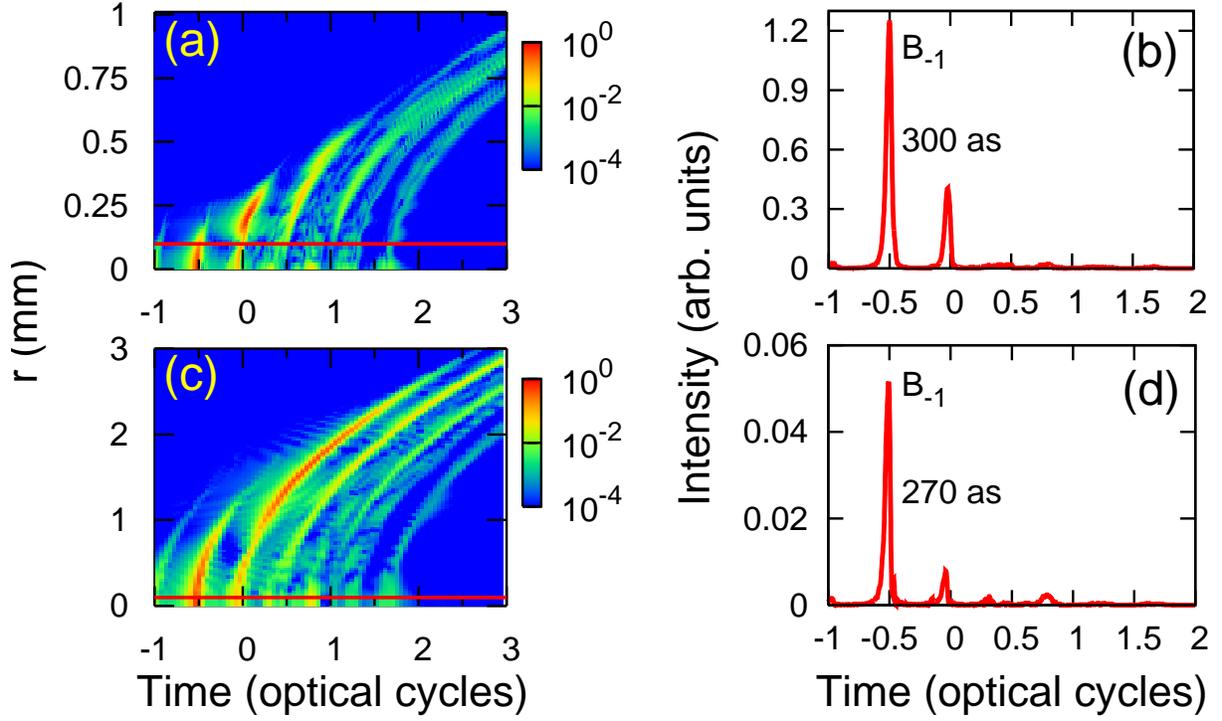}}}}
\caption{(Color online) Dependence of attosecond pulse generation on
the filter position in the far field. Spatial distribution [(a) and
(c)] (normalized) and attosecond pulses synthesized [(b) and (d)] at
different far-field positions: z=100 mm (top row), and z=900 mm
(bottom row), using H40-H80. Laser intensity is 1.0$\times$10$^{14}$
W/cm$^{2}$. These figures are to be compared to Figs.~\ref{Fig4}(e)
and (f) for z=455 mm. Intensity of attosecond pulses in the far
field is obtained using a spatial filter with a fixed radius
r$_{0}$=100 $\mu$m [indicated by the solid line in red in (a) and
(c)]. The main emission burst is labeled with the electron born time
t=-1, or B$_{-1}$. The calculation is for CEP=0. \label{Fig6}}
\end{figure*}

The position of the spatial filter in the far field can be easily
adjusted in an experiment. Here we show the change of attosecond
pulses with the far-field position. In Fig.~\ref{Fig6}, attosecond
pulses synthesized (H40-H80) at two other positions z=100 mm and 900
mm are given. This is to be compared with the ones at z=455 mm in
Figs.~\ref{Fig4}(e) and (f). Using the same filter (indicated by the
solid line in red for fixed r$_0$=100 $\mu$m), the attosecond pulses
generated are shown in Figs.~\ref{Fig6}(b) and (d). The width of the
main burst does not change much with z, but the strength of the
satellite peak is reduced. Of course this is achieved at the expense
of decreasing the strength of the attosecond pulse. This has also
been shown by Gaarde and Schafer \cite{mette-ol-2006} where an IAP
was selected by moving the reflecting mirror further from the laser
focus.

\subsection{Comparison between QRS and SFA in modeling propagation effect}
\begin{figure*}
\mbox{\rotatebox{0}{\myscaleboxc{
\includegraphics[trim=0mm 0mm 0mm 0mm, clip]{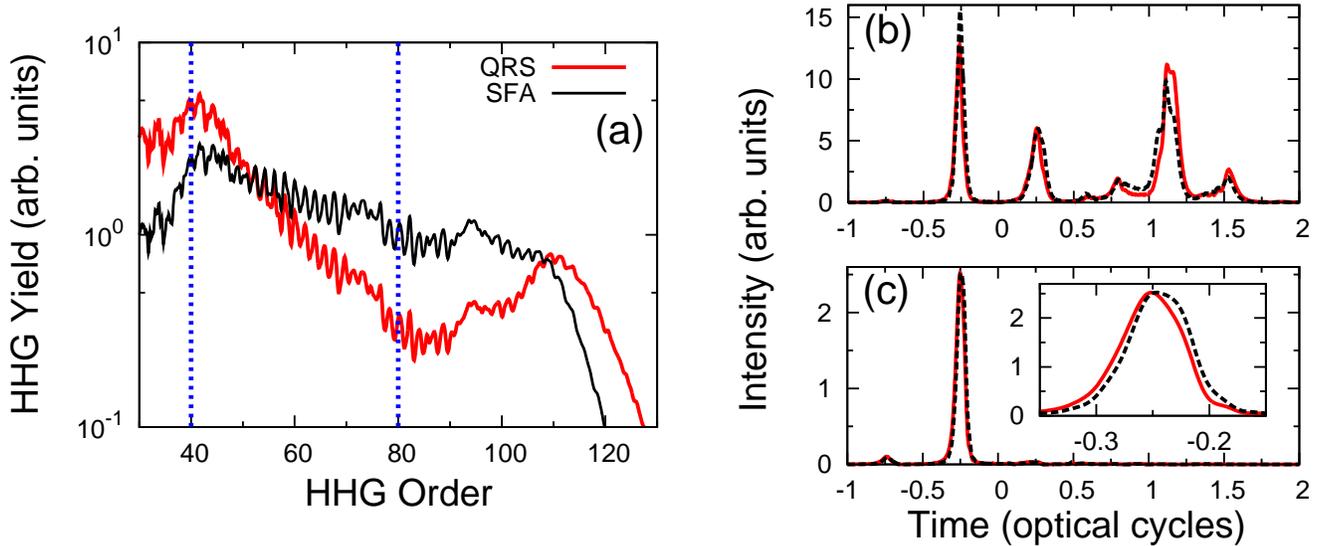}}}}
\caption{(Color online) Comparison of HHG spectra and attosecond
pulses calculated using QRS and SFA for single-atom induced dipoles.
(a) Macroscopic HHG spectra (total spectra without using a slit) of
Xe by QRS [red (dark gray) line] and SFA [black (light gray) line].
Laser parameters: I=1.0$\times$10$^{14}$ W/cm$^{2}$ and CEP=$\pi$/2.
Intensity of attosecond pulses (b) in the near field, and (c) in the
far field (z=455 mm) using a spatial filter with a radius
r$_{0}$=300 $\mu$m: QRS [red (solid) lines] vs SFA [black (dashed)
lines]. Inset in (c): enlarged temporal structure of an IAP. The
spectra are normalized at the peak intensities in (c). The same
normalization factor is used in (a) and (b). H40-H80 are used to
synthesize attosecond pulses.\label{Fig7}}
\end{figure*}

In the last two decades, the strong-field approximation (SFA), which
is in the frame of the SAE approximation, has been widely used to
predict the temporal structure of attosecond pulses even though SFA
is unable to explain the observed harmonic spectra precisely in
general. In the present calculation, we use QRS in the propagation
calculation. For single-atom response, QRS has been tested against
TDSE, both for the magnitude and phase, as documented in Le {\it et
al.} \cite{at-pra-2008}, for example. In QRS, the wave packet is
obtained from SFA, including the phase. The transition dipole
$d(\omega)$ in Eq.~(\ref{single-qrs}), also introduces a phase. In
SFA, this phase is a constant, either real or pure imaginary
(depending on the symmetry of the ground state) and independent of
the harmonic order. In QRS, the transition dipole moment is a
complex number in general. From PI theory, however, it is known that
the phase of the transition dipole does not change much with the
photon energy. Thus the phases of the harmonics calculated from QRS
and SFA do not differ significantly. Since the phases of the
harmonics are much more important in synthesizing attosecond pulses
\cite{mette-prl-2002}, this explains why propagation theory based on
SFA has been so successful in explaining the generation of
attosecond pulses, in spite of its failure in predicting or
explaining the observed harmonic spectra. In this subsection, we
support this analysis with actual results from simulations.

In Fig.~\ref{Fig7}(a) the HHG spectra of Xe obtained from SFA
(within the SAE approximation) and QRS (including multi-electron
effects) using the laser parameters given in the captions are shown.
Clearly the spectra differ greatly. In Figs.~\ref{Fig7}(b) and (c)
the synthesized (H40-H80) attosecond pulses at the near field and
the far field are shown. Clearly the results from the two
calculations are essentially identical (after an overall
normalization), in spite of the large differences in the harmonic
spectra. We have checked some other cases and found that the
temporal structures of the attosecond pulses from the two theories
are always very similar. Larger differences than those shown in
Figs.~\ref{Fig7}(b) and (c) are expected if a wider range of
harmonics are used or if the spectra from the two theories differ
much more, but the general conclusion is correct.

\subsection{CEP dependence of isolated attosecond pulses}
\begin{figure*}
\mbox{\rotatebox{270}{\myscaleboxa{
\includegraphics[trim=-1mm -2mm -4mm 0mm, clip]{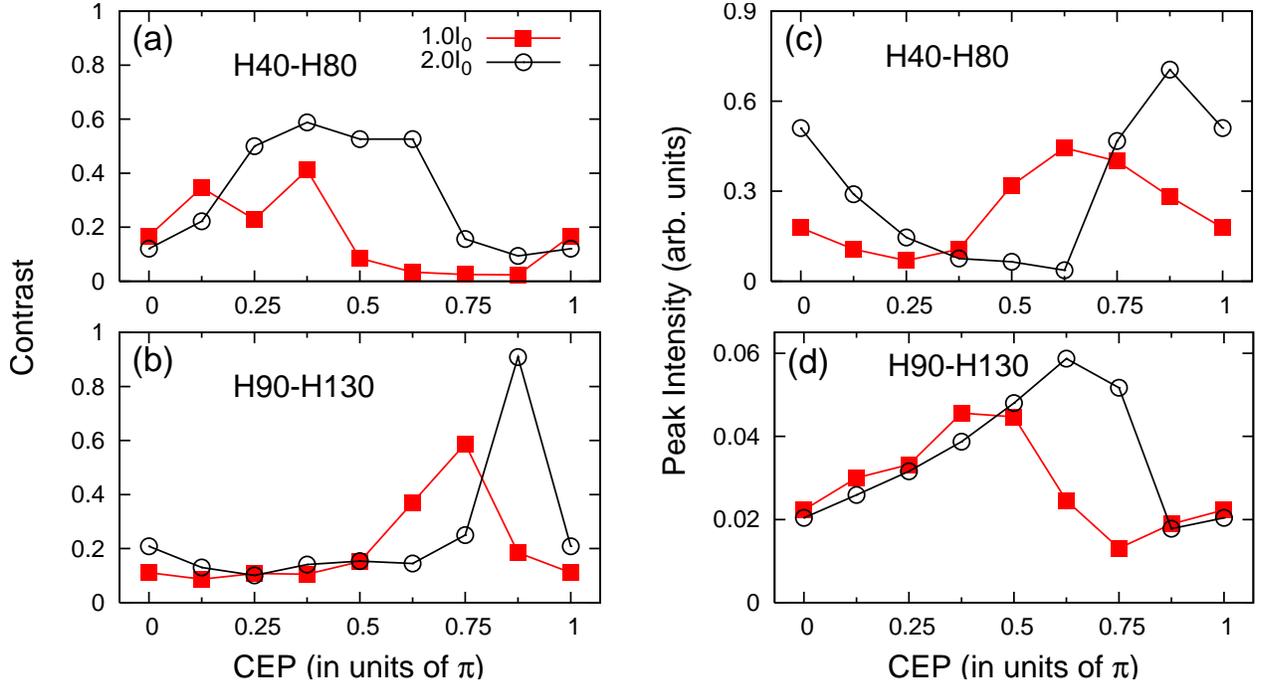}}}}
\caption{(Color online) (a) and (b): The contrast ratio between the
intensities of the strongest satellite and the main attosecond
burst, (c) and (d): The peak intensity of the main attosecond burst
as a function of CEP. Laser intensities are shown in units of
I$_{0}$=10$^{14}$ W/cm$^{2}$. Harmonics used to generate the IAP are
labeled. Far-field position: z=455 mm, and the radius of the spatial
filter: r$_{0}$=100 $\mu$m.\label{Fig8}}
\end{figure*}

The selection of an IAP by a spatial filter in the far field
discussed above is only for a single CEP, and thus only useful if
the laser is CEP-stabilized (has not been achieved for 1.8-$\mu$m
lasers yet). To check if the method can be used for lasers that are
not CEP-stabilized, we investigate the CEP dependence of the IAP
generation.

In Figs.~\ref{Fig8}(a) and (b), we show the contrast ratio between
the intensities of the strongest satellite and the strongest
attosecond burst, and in Figs.~\ref{Fig8}(c) and (d), we show the
peak intensity of the strongest attosecond burst, as the CEP is
varied, for the two laser intensities indicated. A good IAP is to
have high peak intensity for the main peak and weak satellites. From
Figs.~\ref{Fig8}(c) and (d), we note that at the CEP's where the
strongest attosecond bursts have high peak values, the contrast
ratios shown in Figs.~\ref{Fig8}(a) and (b) at these CEP's are
always small. In the meanwhile, when the contrast ratio is large,
the strongest attosecond burst is always weak. Thus it is possible
to generate single attosecond pulses even when the CEP of the
driving laser is not stabilized. This explains the success why the
first single attosecond pulses were generated using few-cycle laser
pulses that were not phase-stabilized \cite{krausz-nature-2001}.

\section{Conclusions}
In this paper, we have studied the generation of isolated attosecond
pulses (IAPs) using few-cycle mid-infrared lasers at large
intensities near and above the critical intensity of Xe. The
calculations are based on the QRS theory where many-electron effects
are included in the single-atom induced dipole moment; specifically,
by including the coupling of the inner 4d shell of Xe in the partial
5p photorecombination transition dipole matrix element. The effect
of the medium on the fundamental and harmonic fields is obtained by
solving the Maxwell's wave equations. The modification (or reshaping
in space and time) of the fundamental field is due to its nonlinear
interaction with the medium includes dispersion, plasma effect, and
Kerr nonlinearity. We have investigated the spatiotemporal evolution
of the fundamental laser field in detail, and found that its
reshaping is responsible for the continuum structure in the HHG
spectra. This conclusion is carried out in terms of the
time-frequency analysis of harmonics in the near and far fields.

Since the divergence of harmonic emission from different half cycles
is varied due to the blue shift and defocusing of the fundamental
laser pulse (or complicated reshaping), we have shown that isolated
attosecond pulses can be generated by synthesizing H40-H80 or
H90-H130, selected by a spatial filter centered on the propagation
axis in the far field. The mechanism of IAP generation in this paper
could be called as ``ionization gating". It works for a loosely
focused laser at high laser intensity (above the critical
intensity), which is reshaped as it propagates through the medium
with a moderate gas pressure. A similar approach has been discussed
by Gaarde {\it et al.} \cite{mette-ol-2006,mette-pra-2006} using a
750-nm laser interacting with 135-Torr Ne gas. We have found that it
is easier to reshape the fundamental field using a long-wavelength
laser with a moderate gas pressure ($\sim$ 30 Torr). The extended
harmonic cutoff of Xe leads to a broad range of harmonics available
for IAP generation. This approach is also different from Ferrari
{\it et al.} \cite{nisoli-natph-2010} where low harmonics ($\sim$ 30
eV, which is equivalent to H40 in this paper) are used to generate
the IAP. In addition, we have discussed the possibilities of
improving the intensity of the IAP, such as changing far-field
position, increasing laser intensity, and increasing gas pressure.
We have shown that the method is very robust and an IAP can be
generated even if the laser CEP is not stabilized.

\section{Acknowledgments}
This work was supported in part by Chemical Sciences, Geosciences
and Biosciences Division, Office of Basic Energy Sciences, Office of
Science, U.S. Department of Energy.

\end{document}